\documentclass[fleqn,usenatbib]{mnras}
\DeclareRobustCommand{\VAN}[3]{#2}
\let\VANthebibliography\thebibliography
\def\thebibliography{\DeclareRobustCommand{\VAN}[3]{##3}\VANthebibliography}

\usepackage{graphicx}	
\usepackage{amsmath}	
\usepackage{amssymb}	

\usepackage{booktabs}
\usepackage{tabularx}
\usepackage{algorithm}
\usepackage{algpseudocode}

\usepackage{url}

\title[Generative AI image amplification]{Amplifying the imaging power of digital sky surveys with space telescopes data and generative AI} 

\author[Erukude \& Shamir]{
Sai Teja Erukude$^{1}$, Lior Shamir$^{1}$\thanks{E-mail: lshamir@mtu.edu}
\\
$^{1}$Kansas State University, Manhattan, KS, 66506, USA\\
}

\date{Accepted xxxx xx xx. Received xxxx xx xx; in original form xxxx xx xx}

\pubyear{2026}

\begin{document}
\label{firstpage}
\pagerange{\pageref{firstpage}--\pageref{lastpage}}
\maketitle

\begin{abstract}

While Digital sky surveys provide excellent throughput of image data and can cover a large footprint, their imaging power is normally inferior to that of space-based telescopes. Space-based telescopes, on the other hand, provide excellent imaging power and can image the deep Universe, but cannot provide the same throughput as advanced ground-based sky surveys. Here, we utilize generative AI to elevate the quality of galaxy images taken by ground-based telescopes to the level of details enabled by space telescopes. The solution is based on the nature of galaxy shapes, allowing generative AI trained on space-based images to convert weak signal into detailed and clear galaxy images. The method allows for combining the high throughput of ground-based sky surveys with the image quality of space-based telescopes. The source code for the method is available, as well as paired training data and a catalog of 63,202 galaxy images enhanced by the proposed method. We also provide a software tool that encapsulates the entire pipeline and the custom generative AI model to generate galaxy images with enhanced quality.

\end{abstract}


\section{Introduction}
\label{introduction}

Digital sky surveys have had a transformative impact on astronomy research \citep{kron1995digital,margony1999sloan,djorgovski2001exploration,ivezic2012galactic,tyson2012future}. Powered by robotic telescopes, digital sky surveys image the sky continuously, collecting and storing image data. These data can be accessed by the public through the concept of virtual observatory. 

Sky surveys such as the Panoramic Survey Telescope and Rapid Response System (Pan-STARRS) \citep{kaiser2002pan}, the Hyper Suprime-Cam (HSC) \citep{aihara2018hyper}, Sloan Digital Sky Survey (SDSS) \citep{york2000sloan}, the Vera Rubin Observatory \citep{ivezic2019lsst}, and the Dark Energy Survey (DES) \citep{dark2016dark} continuously image the sky and collect extremely large astronomical data. 

Another revolutionary astronomical research instrument is the space telescope. Space telescopes such as the Hubble Space Telescope (HST), the James Webb Space Telescope (JWST), Euclid \citep{mellier2024euclid}, and Roman \citep{spergel2015wide}, have been providing image data of astronomical objects with unprecedented quality to transform our understanding of the Universe. 

While the imaging power of space-based telescopes cannot be met by Earth-based telescopes, the throughput of space-based telescopes is lower compared to Earth-based digital sky surveys such as the Vera Rubin Observatory. Therefore, an ideal astronomical imaging device would be a combination of the throughput and sky coverage of Earth-based digital sky surveys, with the image quality of space-based telescopes. Such a system will allow for imaging a large number of objects while also providing the ability to analyze their shape \citep{shamir2009automatic,dieleman2015rotation,semenov2025galaxy,cecotti2020rotation,erukude2025galaxy,elfattah2012automated}.

Here, we use the concept of generative AI to elevate the image quality of Earth-based telescopes to the level of space-based telescopes. Generative AI has been used in image computing for numerous purposes, ranging from the generation of ``deep fake" images to the creation of computer art. Among other tasks, it allows for generating a synthetic image based on information learned directly from the target images. The concept of generative AI has also been used to generate synthetic galaxy images  \citep{campagne2025galaxy,fussell2019forging,lanusse2021deep}, or improving the resolution of low images \citep{hettiarachchi2024generative}.

By using a very large set of images of the same objects taken by both Earth-based and space-based telescopes, we develop a generative AI model that elevates the quality of the Earth-based images. The model identifies the weak signal in galaxy images and amplifies that signal through a generative AI system that is sensitive to the patterns and the visual content of the object. In that sense, the generative AI is used as a complex filter that enhances weak signal to turn it into clear image details.

That provides a computational solution to the enhancement of galaxy images taken by Earth-based telescopes, with no need for new hardware or optics. Its application to digital sky surveys can combine the footprint and throughput of Earth-based digital sky surveys with the imaging power of space telescopes.


\section{Data}
\label{data}


The proposed solution is based on a Conditional Generative Adversarial Network (cGAN) that elevates a ground-based image to a quality equivalent to an image taken by a space telescope. Therefore, the data required for training should include paired images, where one image is acquired by a ground-based instrument and the other shows the exact same astronomical object, but taken by a space telescope. Here, the space-based image data are taken from the Hubble Space Telescope (HST), and the ground-based image data are taken by the Dark Energy Spectroscopic Instrument (DESI) Legacy Imaging Survey \citep{dey2019overview}. Although modern digital sky surveys generate very large databases, the requirement to have galaxies imaged by both the space-based telescope and the ground-based telescope reduces the total size of images that can be used for such a training dataset.

The HST images \citep{shamir2021automatic} were taken from the Cosmological Evolution Survey (COSMOS) \citep{scoville2007cosmos}. The sources were detected by applying SExtractor \citep{bertin1996sextractor} and selecting sources with a magnitude of 4$\sigma$ or higher compared to the background. The sources were then separated by using the Subimage tool of Montage \citep{berriman2004montage}. The images were FITS format images of dimensionality 122 $\times$ 122 pixels, and these images were converted to the simpler TIF format with a dynamic range of 16-bit for the image processing. The entire dataset contained 20,000 objects.  

Each HST image was paired with an image taken from the DESI Legacy Survey centered at the same coordinates. That was done by retrieving image data from the DESI Legacy Survey at the same coordinates as the COSMOS images. Images were retrieved in both the JPG and FITS formats using the cutout API of the ninth data release (DR9). The JPG format has the advantage of size and is also a highly common format used as input for GAN models. On the other hand, the FITS format provides much more detailed information that will be needed to reconstruct the shape of the object from a weak signal. While the human eye might not always be sensitive to the high dynamic range enabled by the FITS format, the information can be used by the AI model to identify subtle patterns in the galaxy shapes that can be observed clearly when using space-telescope images. Because the physics of galaxy formation leads to repetitive shapes, the identification of such patterns can enable the cGAN to reconstruct the fine details of the shape of the galaxy. Both formats were tested, which involved using two separate datasets of paired images in two distinct experiments.

\begin{figure}
    \centering
    \includegraphics[scale=0.5]{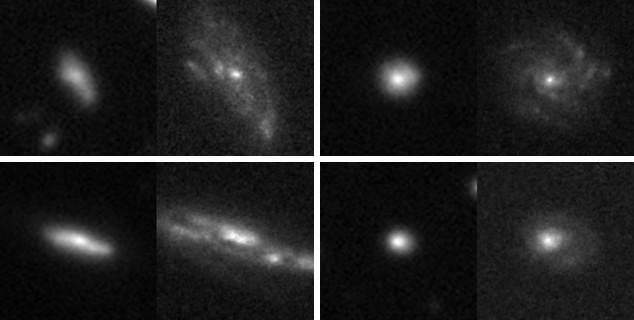}
    \caption{Pairs of training images used to train the cGAN. Each pair includes an image taken by the DESI Legacy Survey (left) and an image of the same object taken by the HST (right).}
    \label{fig_training_samples}
\end{figure}

Figure~\ref{fig_training_samples} shows examples of the galaxy images taken by DESI Legacy Survey and the corresponding galaxy image taken by the HST. As the figure shows, the HST images are far more detailed compared to the images taken by the DESI Legacy Survey, providing more information about the shape of the galaxies that cannot be seen by observing the images acquired by the Earth-based telescope.

\subsection{Data pre-processing pipeline}
\label{preprocessing}

\begin{figure}
    \centering
    \includegraphics[scale=0.31]{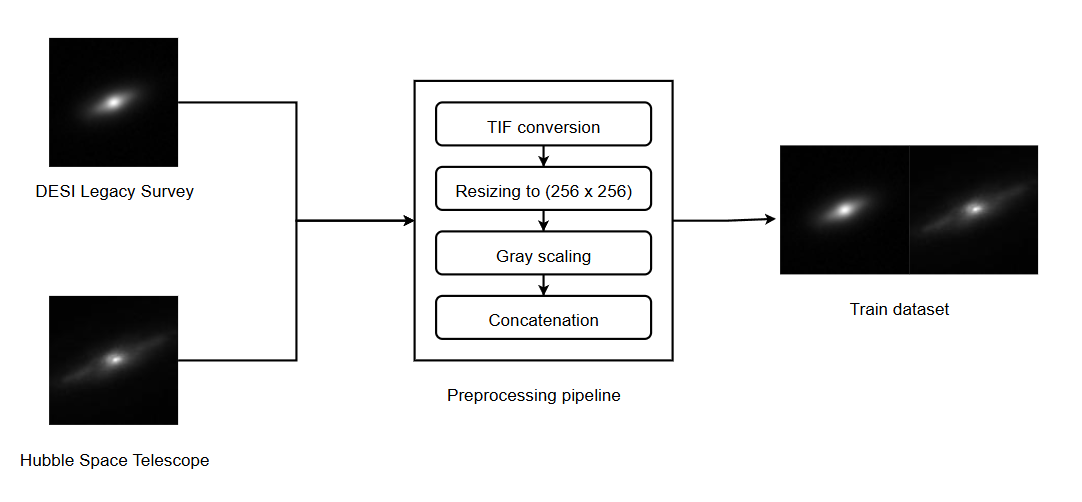}
    \caption{Overview of the data pre-processing pipeline, illustrating the inputs, the output, and key transformation steps involved.}
    \label{fig_preprocessing}
\end{figure}

The DESI Legacy Survey and Hubble Space Telescope images are first passed through a pre-processing pipeline to prepare them for training. This pipeline consists of several key steps, as illustrated in Figure~\ref{fig_preprocessing}. It begins with converting the original FITS files into 16-bit TIF format by averaging pixel values along the first dimension (axis 0) to produce a single 2D grayscale image. The purpose of the conversion to the TIF format was to use a format compatible with the architecture, while the 16-bit dynamic range ensured that subtle pixel value differences are preserved through the format conversion. Next, all images are resized to 256$\times$256 pixels. Finally, each ground-based image is concatenated with its corresponding space telescope image counterpart side by side. This results in a training dataset comprising 20,000 image pairs used to train the cGAN. The training images are available in ``train.zip'' at \url{https://doi.org/10.6084/m9.figshare.30226591}.

\section{A cGAN-based method for amplifying the imaging power of ground-based telescopes}
\label{method}

A Generative Adversarial Network \cite{goodfellow2014generative} is a deep learning concept used to generate synthetic data designed to mimic original data. A GAN is made of two neural networks: the generator and the discriminator. The generator converts input data into synthetic data optimized to be indistinguishable from real data. The discriminator acts like a regulator, attempting to identify the synthetic data from the real data. It tries to classify between the generated sample and the real sample to estimate the differences between them. The generator learns from the discriminator's feedback to produce samples that are more similar to the real samples. This is done iteratively, with the generator and discriminator both improving through the process.

We use the Conditional Generative Adversarial Network (cGAN) to improve the quality of ground-based images. The GAN can learn from the patterns of galaxy shapes in the ground-based telescope images by comparing them to the detailed space-based images of the same objects. By pairing each ground-based image with the space-based image, the GAN learns how to convert weak signal that is difficult to observe by eye into the detailed full shape as observed by a powerful space telescope.

For that purpose, we use the paired cGAN approach of image-to-image conversion \cite{isola2018imagetoimagetranslationconditionaladversarial}.  The networks are trained such that the input pair is an image taken by the ground-based telescope, paired with an image of the same object taken by the space telescope. Through training, the system learns the links between the weaker patterns identified in the Earth-based images and their corresponding detailed shapes in the space telescope images. After the system is trained to identify the links between the paired images, it can enhance new ground-based images and turn them into detailed images of quality typical of space telescopes.

\subsection{cGAN Architecture}

\begin{figure*}
    \centering
    \includegraphics[width=5in]{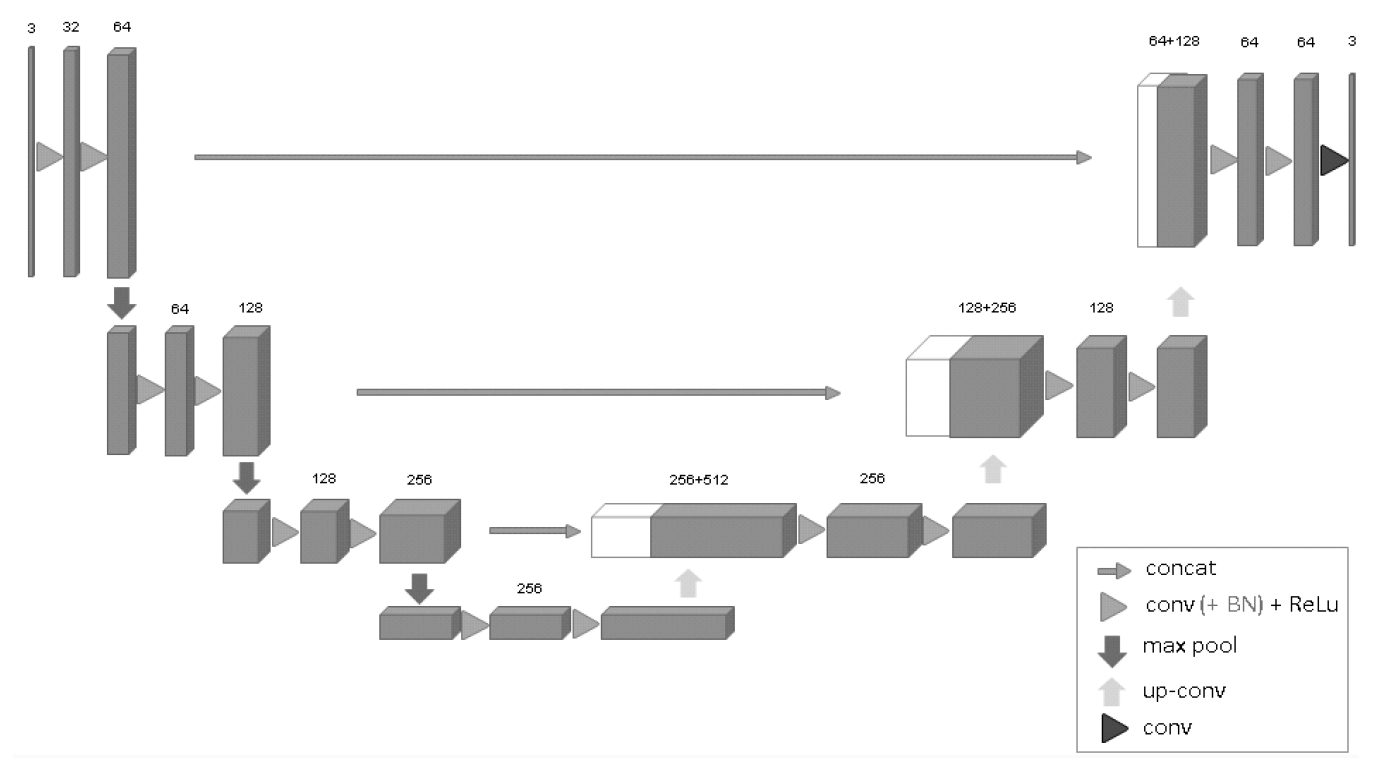}
    \caption{U-Net architecture of the generator used in the enhancement of ground-based telescope images.}
    \label{fig_generator_architecture}
\end{figure*}

The generator is based on the U-Net-based architecture \citep{ronneberger2015u}, which has been found effective also for astronomy images \citep{erukude2025galaxy}. The U-shaped structure is made of a contracting path (encoder) and an expansive path (decoder), with skip connections between the corresponding layers and the encoder and decoder paths, as shown in Figure~\ref{fig_generator_architecture}. This design allows the combination of high-level information with low-level details, and can therefore amplify the visual information into the fine details of space-based telescopes. 

\begin{figure}
    \centering
    \includegraphics[scale=0.5]{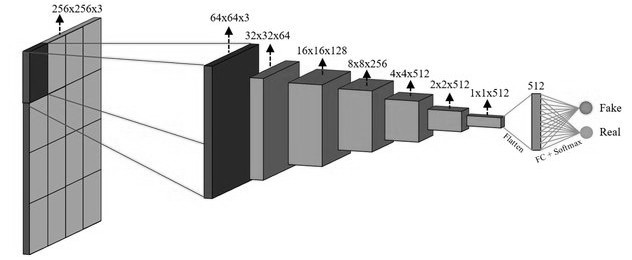}
    \caption{PatchGAN-based architecture of the discriminator.}
    \label{fig_discriminator_architecture}
\end{figure}

The discriminator is based on a PatchGAN architecture \citep{demir2018patch}, as shown in Figure \ref{fig_discriminator_architecture}. The PatchGAN discriminator is trained and classifies N$\times$N patches of the input images rather than the full images. This design is effective for enhancing high-frequency information, helping to preserve the fine details in the generated images. 

The loss $L_g$ is the commonly used weighted sum of the adversarial loss $L_{gan}$ and the $L_1$ loss, as shown by Equation~\ref{loss_function}. $L_{gan}$ is the discriminator loss, and $l_1$ is the generator loss.  $\lambda$ is a hyperparameter set to 0.5.

\begin{equation}
L_g = L_{gan} + \lambda \cdot L_1.
\label{loss_function}
\end{equation}

\subsection{Training the model}

The model was trained using the dataset of image pairs described in Section~\ref{data}. Each training sample is a pair of images, where one image is the ground-based telescope image provided by the DESI Legacy Survey. The other image is the image of the same galaxy captured by the space-based HST, which is the target image that the cGAN is trained to generate automatically from the source image.

We trained the model using the Adam optimizer \citep{kingma2014adam} with a learning rate of 0.0002, a batch size of 1, and a total of 100 epochs. For transparency, all training hyperparameters are listed in Table~\ref{tab_training_config}. Table~\ref{parameters} shows the number of parameters used in each of the models. Unless otherwise noted, we report results using the model checkpoint from epoch 60, as discussed in Section~\ref{results}. All experiments were executed on a high-performance computing system, with the hardware and software details provided in Table~\ref{tab_server_config}. 

\begin{table*}
\small
\caption{Training configuration summary for the proposed Pix2Pix-based generative AI framework.}
\label{tab_training_config}
\begin{tabularx}{\textwidth}{lX}
\toprule
\textbf{Component} & \textbf{Setting} \\
\midrule
Model architecture   & Pix2Pix conditional GAN (U-Net generator, PatchGAN discriminator) \citep{isola2018imagetoimagetranslationconditionaladversarial} \\
Generator backbone   & U-Net encoder C64--C512; decoder CD512--C64 \citep{ronneberger2015u} \\
Discriminator        & PatchGAN, C64--C512, Sigmoid patch output \citep{demir2018patch} \\
Input resolution     & \(256 \times 256 \times 3\) RGB tiles (ground/space pairs) \\
Dataset type         & Paired image translation (Ground $\rightarrow$ Space telescope) \\
Normalization        & Pixel values scaled from [0, 255] to [-1, 1] \\
Batch size           & 1 \\
Optimizer            & Adam, learning rate \(0.0002\), \(\beta_1 = 0.5\) \\
Loss functions       & Adversarial BCE + L1 (MAE) reconstruction \\
Loss weighting       & Generator loss: BCE : L1 = 1 : 100 \\
Training epochs      & 100 \\
Model saving         & Periodic generator checkpoints (every 10 epochs) \\
Implementation framework & TensorFlow / Keras \\
Repository           &  \url{https://github.com/SaiTeja-Erukude/Enhancing-Ground-Based-Astronomy-using-GenAI}\\
\bottomrule
\end{tabularx}
\end{table*}

\begin{table}
\caption{Hardware and software configuration of the experimental server.}
\centering
\begin{tabular}{ll}
\toprule
\textbf{Component}       & \textbf{Specification} \\
\midrule
Operating System         & Debian GNU/Linux 12 (bookworm) \\
Kernel Version           & 6.1.0-28-amd64 \\
CPU Model                & 2 × Intel Xeon Gold 5317, 3.0 GHz \\
CPU Cores / Threads      & 48 cores / 96 threads \\
System Memory            & 125 GiB RAM \\
Primary Storage          & 223 GB SSD (system) \\
Secondary Storage        & 5.2 TB HDD (data) \\
\bottomrule
\end{tabular}
\label{tab_server_config}
\end{table}

\begin{table}
\caption{Number of parameters in the models.}
\centering
\begin{tabular}{lccc}
\toprule
\textbf{Model}       & \textbf{Trainable} & \textbf{Non-trainable} & \textbf{Total} \\
      & \textbf{parameters} & \textbf{parameters} & \textbf{parameters} \\
\midrule
Generator  & 54,419,459 & 9,856  & 54,429,315 \\
/ U-Net  & & & \\
Discriminator   & 6,965,441 & 2,816 & 6,968,257 \\
/ PatchGAN  & & & \\
Full cGAN & 54,422,275	& 6,975,297 & 61,397,572 \\
\bottomrule
\end{tabular}
\label{parameters}
\end{table}

\subsection{Algorithms}

The complete training and inference procedures of the proposed generative AI framework are summarized using two stage-wise algorithms. Algorithm~\ref{algo_training} describes the standalone training procedure for the Pix2Pix Conditional GAN model that amplifies the imaging power, while Algorithm~\ref{algo_inference} outlines the inference pipeline used to generate the space-quality enhanced outputs from ground-telescope images.

\begin{algorithm}
\caption{Training procedure for the Pix2Pix conditional GAN model.}
\label{algo_training}
\begin{algorithmic}[1]
\Require \\
    Ground–space paired dataset $(A,B)$ where:  $A =$ ground–telescope images (input domain) and $B =$ space–telescope images (target domain) \\
    Discriminator $D$, Generator $G$, Adversarial composite model $GAN$ \\
    Epochs $E$, Batch size $b$

\State Compute number of batches $N \leftarrow \lfloor |A|/b \rfloor$
\State Compute total update steps $T \leftarrow E \cdot N$

\For{$t = 1$ to $T$}
    \State Select real sample pairs $(x_A, x_B)$ from dataset
    \State Compute real labels $y_{\text{real}}$
    \State Generate fake target samples:
    \[
        \hat{x}_B \leftarrow G(x_A)
    \]
    \State Compute fake labels $y_{\text{fake}}$
    \State Update discriminator on real:
    \[
        \mathcal{L}_{D1} \leftarrow D.\text{train\_on\_batch}([x_A,x_B], y_{\text{real}})
    \]
    \State Update discriminator on fake:
    \[
        \mathcal{L}_{D2} \leftarrow D.\text{train\_on\_batch}([x_A,\hat{x}_B], y_{\text{fake}})
    \]
    \State Update generator using composite objective:
    \[
       \mathcal{L}_{G} \leftarrow GAN.\text{train\_on\_batch}(x_A, [y_{\text{real}},x_B])
    \]
    \If{$t \bmod (10 \cdot N) = 0$}
        \State Evaluate model on held-out samples and store results
        \State Save the trained generator model $G^\ast$
    \EndIf
\EndFor
\end{algorithmic}
\end{algorithm}

\begin{algorithm}[ht]
\caption{Inference Procedure for Enhancing Ground Telescope Images}
\label{algo_inference}
\begin{algorithmic}[1]
\Require  \\
Trained generator $G^\ast$ \\
input image $x_A$ from domain $A$ (ground–telescope image)

\State Preprocess $x_A$ (resize, grayscale conversion, normalization)

\State Generate enhanced output:
\[
    \hat{x}_B \leftarrow G^\ast(x_A)
\]
\Comment{$\hat{x}_B$ approximates the target domain $B$: space–telescope image}

\State Postprocess $\hat{x}_B$ to valid pixel range
\State \Return enhanced image $\hat{x}_B$ (space–quality reconstruction)
\end{algorithmic}
\end{algorithm}

\section{Results}
\label{results}


Evaluation of the effectiveness of GANs is known to be a challenging task \cite{shmelkov2018good,borji2022pros}. GANs are often used to generate complex data, making it difficult to assess the quality of the output quantitatively. In some cases, cognitive tests are used to determine the ability of the GAN to generate images that seem to a person similar to natural images, or other target images that the GAN is expected to generate \citep{arora2021review,ben2024overview,zhang2017style,lang2021explaining}. The analysis of the GAN is performed here using the loss function, as well as through manual inspection by comparing the generated images to the same images captured by the Hubble Space Telescope. Another method of evaluation is applying basic galaxy image analysis on the galaxy images acquired by the space telescope and the AI-generated galaxy images, and comparing the differences between the results.

\begin{figure}
    \centering
    \includegraphics[scale=0.5]{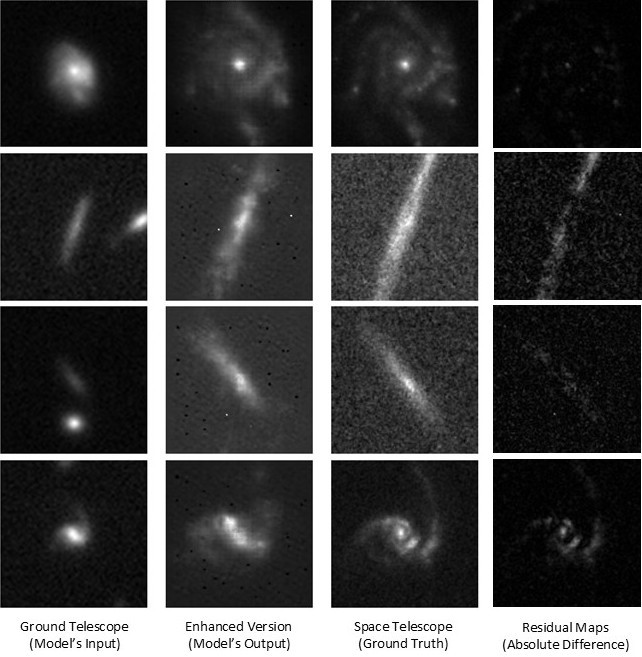}
    \caption{Examples of ground-based telescope images, the space-based images of the same objects, and the images generated by the GAN from the ground-based images. The residual maps are created from the differences between generated images and the space telescope images. The objects are at coordinates $(\alpha=150.6532946^o,\delta=1.6253657^o)$, $(\alpha=150.3196276^o,\delta=1.7540747^o)$, $(\alpha=150.3742923^o,\delta=1.6191228^o)$, $(\alpha=150.0445515^o,\delta=1.6114927^o)$.}
    \label{fig_catalog_samples}
\end{figure}

Figure~\ref{fig_catalog_samples} shows examples of DESI Legacy Survey images, and images generated by the GAN when these images are used as input. The generated images can be compared to the images of the same objects taken using the Hubble Space Telescope. As the figure shows, the GAN is capable of reconstructing the morphological features of the galaxy, making the enhanced galaxy similar in detail to the image taken by HST.

While the figure shows just a few examples, these examples are representative of the entire catalog. The entire catalog is available online and described in Section~\ref{catalog}.


\begin{figure}
    \centering
    \includegraphics[scale=0.5]{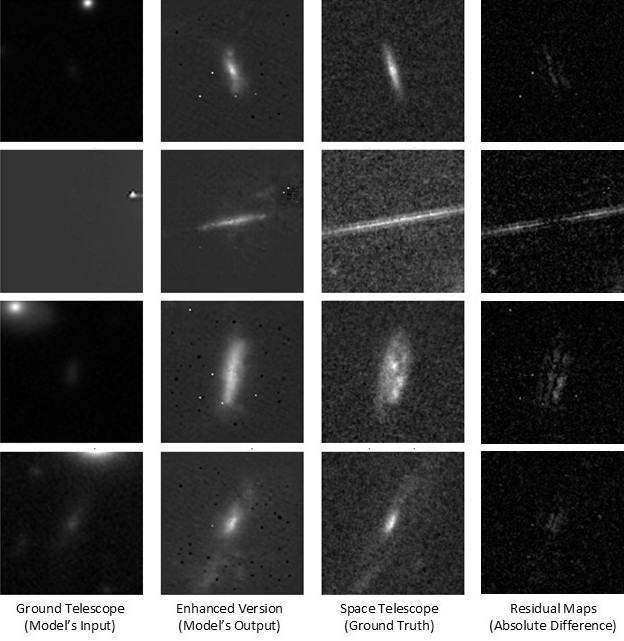}
    \caption{Examples from the catalog where the model demonstrates outstanding performance. Even when the input image lacks visible details of the galaxy, the model generates an enhanced version that closely matches the ground truth. The residual maps also show the good agreement between the generated image and space telescope image. The objects are at coordinates $(\alpha=149.68237^o,\delta=2.4202963^o)$, $(\alpha=149.97054^o,\delta=2.6837897^o)$, $(\alpha=149.8200861^o,\delta=2.4452223^o)$, $(\alpha=149.8254184^o,\delta=2.1643785^o)$.}
    \label{fig_catalog_magical_samples}
\end{figure}

\begin{figure}
    \centering
    \includegraphics[scale=0.5]{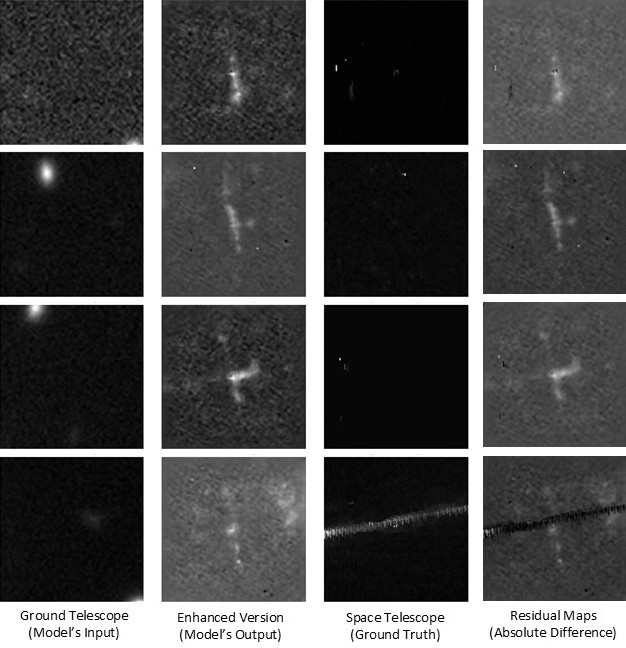}
    \caption{Catalog examples highlighting cases where the model underperforms. These issues might be addressed by training the model on a larger and more diverse set of images. The objects are at coordinates $(\alpha=150.7538768^o,\delta=1.957399^o)$, $(\alpha=150.7714556^o,\delta=2.0616442^o)$, $(\alpha=150.7619236^o,\delta=2.6261117^o)$, $(\alpha=150.7552252^o,\delta=2.2725501^o)$.}
    \label{fig_catalog_bad_samples}
\end{figure}

As Figures~\ref{fig_catalog_magical_samples} and~\ref{fig_catalog_bad_samples} show, the enhanced images also contain some background ``pepper noise" added to the images, which is typical to AI-generated images. That can be removed from the images by using a final step of pre-processing. For instance, a simple low-pass filter can smooth that noise. That, however, can also change some of the relevant visual content. Therefore, removing the ``pepper noise" should be done under the awareness that visual content can be affected by applying such filters.

Another test was done by applying automatic image analysis to the galaxy images. The analysis was first applied to the space-telescopes images, and then the results of each image were compared to the results when applying the same analysis to the AI-enhanced galaxy image. If the AI-enhanced galaxy images are the same as the space telescope images, the results are expected to be identical. Therefore, any difference between the analysis of the space telescope image and the corresponding AI-generated image shows that the enhanced images are different.

For that purpose we used the mature and commonly used SExtractor \citep{bertin1996sextractor}. SExtractor is a software tool for basic analysis of astronomical images. Since it is designed for both point sources and extended sources, it can also be applied to galaxy images. Here we used 1,000 images of galaxies taken by HST, paired with AI enhanced galaxy images of the same galaxies. These pairs of images are taken from the catalog described in Section~\ref{catalog}. 

The analysis included several descriptors computed by SExtractor \citep{bertin1996sextractor}, which are the major axis, minor axis, position angle, elongation, and ellipticity. Additionally, we used the Ganalyzer tool \citep{shamir2011ganalyzer} to determine whether the galaxy is elliptical or spiral, and compare the results computed on the space telescope images to the results computed on the AI-enhanced images. The analysis was done such that elliptical shape was assigned 1, and spiral shape was assigned the value 0.

Table~\ref{sextractor} shows the average relative difference $\frac{\Delta}{\mu}$ between the descriptors computed from the space telescope images and the descriptors computed from the corresponding AI-enhanced images of the same galaxies. As the table shows, the descriptors as measured on the AI-enhanced images are not identical to those determined by the space telescope images, but normally stay within the 5\% difference. The broad morphology reflects 45 galaxies out of the 1,000 that were tested in which the morphology of the AI-enhanced image did not match the morphology determined by using the space telescope image.  

\begin{table}
\small
\caption{The differences between galaxy descriptors computed from the space telescope images and the same attributed computed from the AI-enhanced images of the same galaxies.}
\label{sextractor}
\begin{tabular}{lcc}
\toprule
Descriptors &  & $\overline{\text{Relative difference}}$ \\
\midrule
Major axis (A) &       &      0.056               \\
Minor axis (B) &       &      0.052               \\
Position angle ($\theta$) &       &     0.031     \\
Elongation (A/B) &       &          0.05        \\
Ellipticity 1-(B/A)&       &        0.05          \\
Broad morphology    &        &      0.06              \\
\bottomrule
\end{tabular}
\end{table}

\subsection{A catalog of enhanced galaxy images from DESI Legacy Survey}
\label{catalog}

To test the ability of the method to provide catalogs of enhanced galaxy images, the method was applied to 63,202 galaxies from the DESI Legacy Survey. The galaxies were imaged by the Dark Energy Camera (DECam) of the Blanco telescope in Cerro Tololo, Chile. The catalog is available for download at \url{https://doi.org/10.6084/m9.figshare.30226591}.

\begin{figure*}
    \centering
    \includegraphics[scale=1.0]{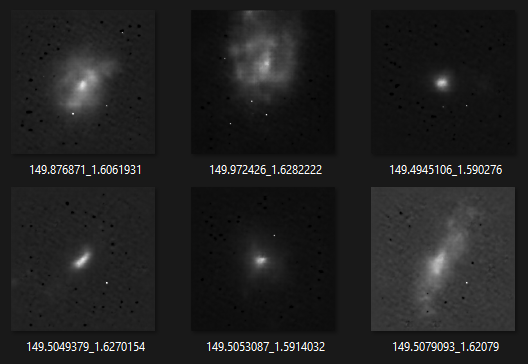}
    \caption{Sample enhanced images from the catalog of enhanced galaxy images.}
    \label{fig_catalog_enhanced_galaxies}
\end{figure*}

The catalog is organized into two directories: ``enhanced\_galaxies'' and ``galaxy\_comparisons''. The ``enhanced\_galaxies'' folder contains the output images produced by our generative AI model, each with a resolution of 256$\times$256 pixels, as illustrated in Figure~\ref{fig_catalog_enhanced_galaxies}.

\begin{figure*}
    \centering
    \includegraphics[scale=1.0]{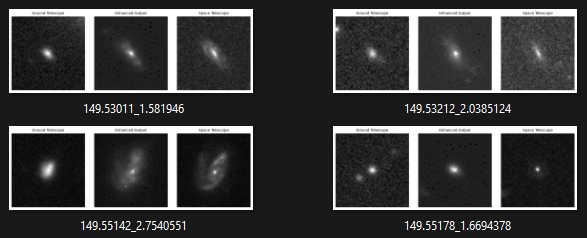}
    \caption{Example images from the ``galaxy\_comparisons" of the catalog.}
    \label{fig_catalog_comparisons}
\end{figure*}

The ``galaxy\_comparisons'' folder includes image collages that provide a visual qualitative assessment of the model's performance. Each collage consists of three images: the original input from a ground-based telescope (DESI), the AI-enhanced output, and the corresponding target image captured by a space-based telescope (ground-truth). These side-by-side comparisons help evaluate how closely the generated images resemble the true observations, as depicted in Figure~\ref{fig_catalog_comparisons}.

To reduce the overall size of the catalog, all images in the catalog are saved in PNG format. To be able to provide the ``ground truth" for each enhanced galaxy image, the catalog is based on galaxies from the footprint covered by the COSMOS field.

Each galaxy in the catalog is identified by the equatorial celestial coordinates, which are embedded in the filenames. These filenames consist of two components separated by an underscore: the first represents the Right Ascension (RA), and the second denotes the Declination (Dec) of the galaxy. For instance, an enhanced galaxy image in the catalog could be named ``150.3295405\_1.6032845.png''. In this case, the coordinates are $(\alpha=150.3295405^o,\delta=1.6032845^o)$.

\subsection{Galaxy Enhancer Software}
\label{galaxy_enhancer}

\begin{figure*}
    \centering
    \includegraphics[scale=0.75]{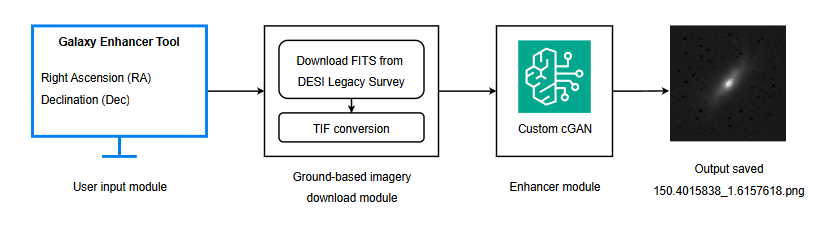}
    \caption{Modules and data flow involved in the galaxy enhancer tool.}
    \label{fig_galaxy_enhancer_tool}
\end{figure*}

To simplify the user experience, we have bundled multiple components into a unified tool called ``Galaxy Enhancer''. As depicted in Figure~\ref{fig_galaxy_enhancer_tool}, this tool is composed of several integrated modules: the ``User Input Module'', the ``Ground-Based Imagery Download Module'', and the ``Enhancer Module''.

The user input module prompts users to provide the celestial coordinates of their target galaxy, specifically, the Right Ascension (RA) and Declination (Dec). 
The corresponding galaxy image is then retrieved in FITS format. 
After downloading, the FITS file is converted to a 16-bit TIF image.

This TIF image is subsequently processed by the Enhancer Module, which houses our pre-trained custom generative AI model, 
as described in Section~\ref{method}. The module processes the input, generates the enhanced output, saves the result, and returns the path to the enhanced image. The output file is typically named using the input coordinates, in the format: ``\{RA\}\_\{Dec\}.png''.

\begin{figure}
    \centering
    \includegraphics[scale=0.55]{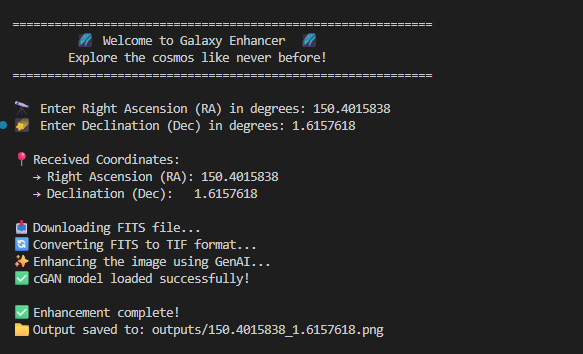}
    \caption{Terminal output illustrating each stage of the galaxy enhancer execution.}
    \label{fig_galaxy_enhancer_output}
\end{figure}

Figure~\ref{fig_galaxy_enhancer_output} demonstrates how the Galaxy Enhancer tool operates within a terminal interface. The full code base for downloading and using this tool is publicly accessible at \url{https://github.com/SaiTeja-Erukude/Enhancing-Ground-Based-Astronomy-using-GenAI/tree/main/galaxy_enhancer}.

\subsection{Experiments with SDSS and JWST galaxies}
\label{sdss_jwst}

\begin{figure}
    \centering
    \includegraphics[scale=0.25]{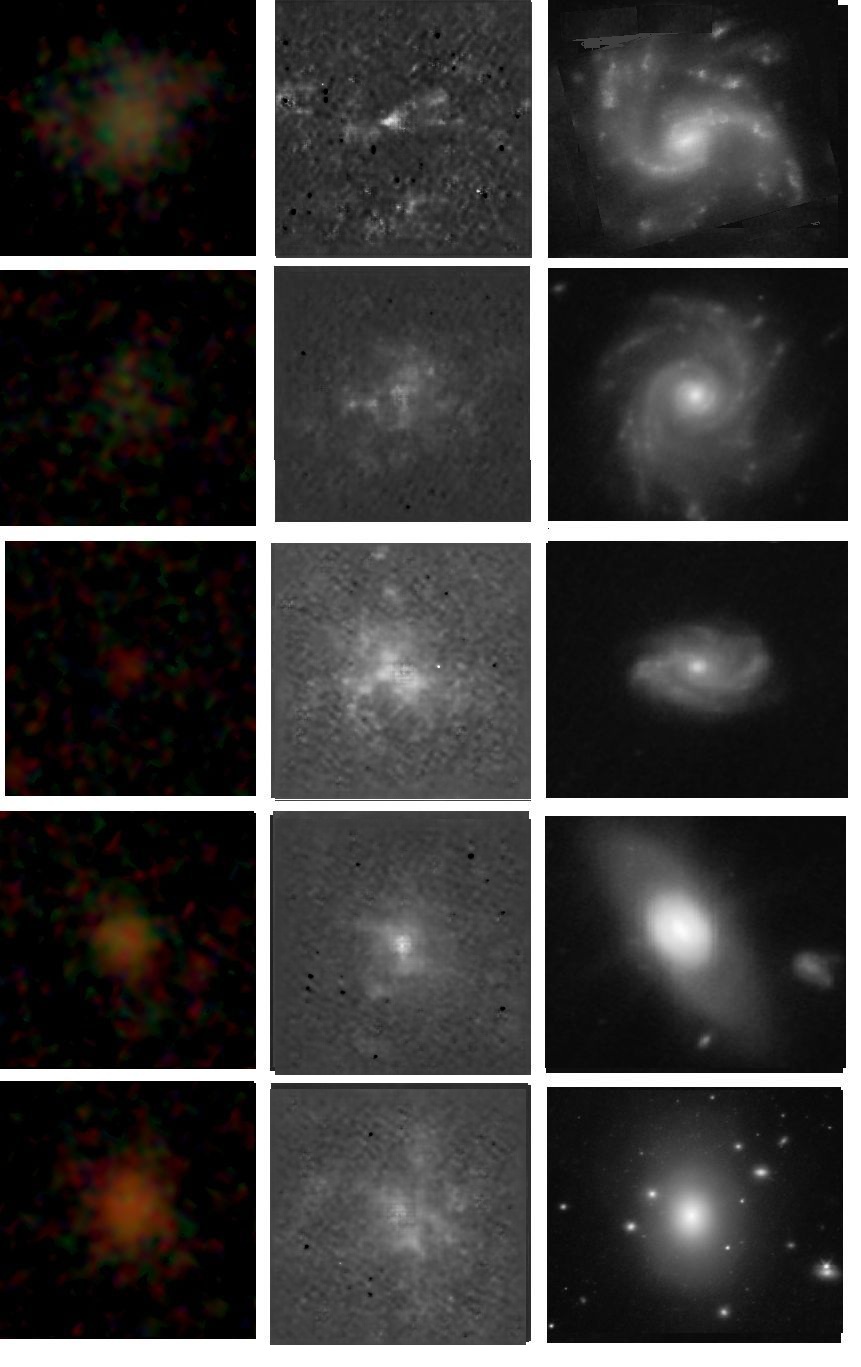}
    \caption{Examples of enhanced SDSS galaxies in comparison to JWST deep field.}
    \label{sdss_jwst_bw}
\end{figure}

Another experiment was done by enhancing galaxies imaged by SDSS, and comparing the results to images taken by JWST deep field. Figure~\ref{sdss_jwst_bw} shows five galaxies imaged by SDSS, the enhanced galaxies, and their comparison to JWST imaged. The JWST field that was used for this experiment is the Stephan's Quintet, at (RA=22h 35m 57.49s, Dec=33$^o$ 57' 36"). The field is highly detailed, with over 150 million pixels, made from almost 1,000 separate images.
Is is also within the footprint of SDSS, allowing to compare the two instruments.

As the figure shows, the AI-enhanced SDSS images do not fit the JWST images at the same quality as the DESI Legacy Survey images enhanced to the quality of COSMOS images. That can be explained by the relatively low details of the SDSS images, compared to the very high level of details of JWST. While the AI can enhance images well, it is limited when the image quality gap between the source and target increases. 

The JWST Stephan's Quintet image is obviously of far higher quality compared to HST COSMOS images, while SDSS images or of lower quality compared to DESI Legacy Survey. That high difference in image quality makes it far more challenging for the AI to enhance the images, showing the limitation of the method.

\section{Conclusion}
\label{conclusion}


Autonomous digital sky surveys are among the most powerful and most productive research instruments of our time. Earth-based digital sky surveys have a high bandwidth of data collection, but the quality of the imaging is still not comparable to space-based telescopes.

Here, we used generative AI to enhance the quality of the images taken by Earth-based digital sky surveys. The enhancement is done by using a pix2pix GAN as a comprehensive filter. It is trained by pairs of images of the same galaxies taken by both Earth-based and space-based telescopes. The GAN is then trained to bridge between the images and can then transform images taken by Earth-based telescopes into the quality typical of space-based telescopes. In that sense, it can transform digital sky surveys into much more powerful instruments, combining their ability to cover large parts of the sky with the imaging quality of space-based telescopes. 

The method was applied to a catalog of a large number of galaxies, demonstrating that it can be used to generate catalogs of enhanced images. It is also provided in the form of a software tool that can transform input galaxy images, and therefore can be used for on-the-fly transformation of the images. The ability to transform the images in real time can be used by digital sky surveys to provide users with the ability to enhance objects of their choice as they browse through the user interface.

While the method can provide additional power to the digital sky survey, it has several limitations. Firstly, it is trained and tested on extended objects only, as point sources are not part of this study. The enhancement is based on repetitive patterns of galaxy shapes, and in some cases of rare objects, the enhancement might lead to an image that is different from what that rare object is. In any case, machine learning systems are normally expected to have a certain degree of errors, and therefore some galaxies might be transformed in a manner that is not consistent with the true visual appearance of the object.

But despite the limitations, the method can be used to enhance the quality of Earth-based images without the need to make the substantial resource investment typical of space-based telescopes. It is also fast and can therefore be used for on-the-fly enhancement of images without necessarily generating dedicated catalogs. Due to its availability and low footprint, such a method can be added to existing and future digital sky surveys to maximize their discovery power.

\section*{Acknowledgments}

We would like to thank the knowledgeable anonymous reviewer for the helpful comments.

\subsection*{Funding}

The research was supported in part by NSF grant number OIA-2148878.

\subsection*{Conflicts of Interest}

The authors declare that there is no conflict of interest regarding the publication of this article.

\subsection*{Data Availability}

The catalog of enhanced galaxy images, as well as training data for the cGAN, can be downloaded at \url{https://doi.org/10.6084/m9.figshare.30226591}. Code used in this project is available at \url{https://github.com/SaiTeja-Erukude/Enhancing-Ground-Based-Astronomy-using-GenAI}.


\bibliographystyle{apalike}
\bibliography{main}

\bsp	
\label{lastpage}
\end{document}